\documentclass[12pt]{article}
\usepackage{a4wide}
\usepackage[utf8]{inputenc}
\usepackage{csvsimple}
\usepackage{amsmath}
\usepackage{amssymb}
\usepackage{cite}
\usepackage{hyperref}
\usepackage{graphicx}
\graphicspath{{./images/}}
\usepackage{float}
\usepackage{booktabs}
\usepackage{bigstrut}
\usepackage{geometry}
\usepackage{fancyhdr}
\usepackage{physics}
\usepackage{mathrsfs} 
\usepackage{bm}
\usepackage{calligra}

\usepackage[dvipsnames]{xcolor}

\usepackage{subfig}

\def\o{\omega}

\def\f{\frac}

\def\mc{\mathcal}

\DeclareMathAlphabet{\mathcalligra}{T1}{calligra}{m}{n}
\DeclareFontShape{T1}{calligra}{m}{n}{<->s*[2.2]callig15}{}
\newcommand{\scripty}[1]{\ensuremath{\mathcalligra{#1}}}

\begin{document}

{\pagestyle{empty}
\pagestyle{plain}
{\renewcommand{\thefootnote}{\fnsymbol{footnote}}
\begin{center}
\vfill
{\Large \bf  Probing the curvature of the cosmos from \\quantum entanglement due to gravity}

\vskip1cm
Suddhasattwa Brahma$^{1}$ and Abhinove Nagarajan Seenivasan$^{2,3}$
\\
\vspace{0.5em}
$^{1}$Higgs Centre for Theoretical Physics, School of Physics and Astronomy, \\ University of Edinburgh, Edinburgh EH9 3FD, United Kingdom\\
$^{2}$Department of Physics, Indian Institute of Technology Guwahati,\\ Guwahati 781039, Assam, India. \\ 
$^{3}$School of Mathematics and Statistics, University of Sheffield, \\ Hicks Building, Hounsfeld Road, Sheffield S3 7RH, United Kingdom \\
\vspace{3.5em}
\end{center}
}

\setcounter{footnote}{0}


\begin{abstract}
	If gravity is fundamentally quantum, any two quantum particles must get entangled with each other due to their mutual interaction through gravity. This phenomenon, dubbed gravity-mediated entanglement, has led to recent efforts of detecting perturbative quantum gravity in table-top experimental setups. In this paper, we generalize this to imagine two idealized massive oscillators, in their ground state, which get entangled due to gravity in an expanding universe, and find that the curvature of the background spacetime leaves its imprints on the resulting entanglement profile. Thus, detecting gravity-mediated entanglement from cosmological observations will open up an exciting new avenue of measuring the local expansion rate of the cosmos.
\end{abstract}

\setcounter{page}{1}

\vfill

\vspace{2mm}
\section{Introduction}
Quantum mechanics and gravity usually do not go together since we are yet to formulate a consistent theory of quantum gravity. However, recent results from black hole physics, in particular, have shown how quantum information theory constitutes an increasingly powerful approach to study problems at the intersection of quantum mechanics and general relativity \cite{Chapman:2021jbh}. On the other hand, as Feynman had wisely prophesied  \cite{DeWitt:1957obj}, creative thought experiments often fill the lacuna in advancing this topic in which we are not pushed by direct experimental evidence \cite{Mari:2015qva, Dyson:2014qoa, Parikh:2020nrd, Bose:2017nin, Marletto:2017kzi, Belenchia:2018szb}. 

It is thus not surprising that gravity-mediated entanglement (GME), which is based on applying modern quantum informatic tools to an old {\it Gedankenexperiment} \cite{DeWitt:1957obj} for testing the quantum nature of the gravitational field, has emerged as one of the most exciting avenues in this area (see, for instance, \cite{Marletto:2018lsb, Bose:2022uxe, Anastopoulos:2015zta, Danielson:2021egj, Biswas:2022qto, Christodoulou:2018cmk, Danielson:2022tdw, Christodoulou:2022knr, doi:10.1142/S0218271823420208}). Such has been the remarkable development in quantum technology over the last few decades that new proposals to detect such GME in a table-top experiment has recently been made (see \cite{Yant:2023smr, Marshman:2019sne, PhysRevLett.127.123602, PhysRevLett.130.240203} and references therein). Such an experiment, if successful, would be a major achievement in proving perturbative quantum gravity inasmuch that gravitational interactions are indeed mediated by gravitons. 
\vspace{2mm}

\section{From flat space to an expanding universe}
 Let us first review the setting of a GME-based experiment in flat space since it would make crossing over to an expanding universe easier. The main idea is that starting with two unentangled particles which are in a product state, if there is a truly quantum (gravitational) field mediating the interaction between them, this would generate an entanglement which can, in principle, be measured. Consider two massive objects whose centres of mass are separated by a distance $d$, at equilibrium, are trapped in one-dimensional harmonic potentials and prepared in their ground states. These identical oscillators, of masses $m$ and frequencies $\omega$ are non-interacting to begin with, and are assumed to be mutually at rest with each other. The free Hamiltonian for this system, at some initial time, is given by
\begin{equation}\label{freeH}
	\hat{H}_0 \equiv \hat{H}^a_0 + \hat{H}^b_0 = \frac{\hat{p}_a^2}{2m} + \frac{1}{2} m \omega^2 \hat{r}_a^2 + \frac{\hat{p}_b^2}{2m} +  \frac{1}{2} m \omega^2 \hat{r}_b^2\,,
\end{equation}
where $\hat{r}_{a,b}$ denotes the displacement of each oscillator from their respective (equilibrium) trap locations such that the distance between the oscillators is given by $\scripty{r}=|d - r_a + r_b|$.  The deviations from the centre of mass position and their conjugate momenta can be written in terms of standard creation and annihilation operators $(\hat{a}, \hat{a}^{\dagger})$ and $(\hat{b}, \hat{b}^{\dagger})$, in the energy basis of $\hat{H}_a$ and $\hat{H}_b$ respectively, as: 
\begin{align}\label{QO}
	\hat{r}_a = \sqrt{\f{\hbar}{2m\o}}\left(\hat{a} + \hat{a}^{\dagger}\right) &~,~~~~ \hat{r}_b = \sqrt{\f{\hbar}{2m\o}}\left(\hat{b} + \hat{b}^{\dagger}\right)~, \\
	\hat{p}_a = i\sqrt{\f{\hbar m\o}{2}}\left(\hat{a}^{\dagger} - \hat{a}\right) &~,~~~~ \hat{p}_b = i\sqrt{\f{\hbar m\o}{2}}\left(\hat{b}^{\dagger} - \hat{b}\right)~.
\end{align} 

 In a Minkowski background, assuming a weak field limit, the gravitational interaction is mediated by $\hat{h}_{\mu\nu}$ where the full background metric is $\tilde{g}_{\mu\nu} = \eta^{\rm flat}_{\mu\nu} + \hat{h}_{\mu\nu}$. The hat over the (linearized) gravitational field emphasizes the fact that we are treating the graviton as a (relativistic) quantum field.  The stress energy tensor $\hat{T}^{\mu\nu}_{a,b} \sim m \,\delta^\mu_0\, \delta^\nu_0\, \delta^{(3)}(\hat{r}_{a,b})$, sourced by the quantum oscillators, appears in the interaction Hamiltonian $\int d^3x~ \hat{h}_{\mu\nu} \hat{T}^{\mu\nu}$. One can solve for the graviton field equations, in terms of standard mode expansions with respect to the Minkowski vacuum, for the full gravity plus oscillator system, and plug it back into the on-shell action, to derive the interaction term between the two oscillators. Restricting to the quadratic expansion of the gravitational action, one can compute this exactly only in the non-relativistic (stationary oscillators) and near-field (instantaneous interaction) approximations \cite{Christodoulou:2022mkf} to find \cite{Bose:2022uxe}\footnote{From hereon, we will use natural units $c=\hbar=1$ and drop the hats on the quantum operators.}
\begin{eqnarray}\label{New_flat_pot}
U_{\rm int}^{\rm flat} = - \frac{Gm^2}{\scripty{r}}= - \frac{Gm^2}{|d + r_a - r_b|}\,.
\end{eqnarray}
 
At first sight, this seems like a trivial result since this is merely the interaction energy due to Newtonian gravity. However, since one carries out a fully relativistic field-theoretic treatment of the graviton as a quantum field, the above line of argument successfully derives the non-relativistic approximation to the scattering amplitude due to the exchange of off-shell gravitons between the two oscillators \cite{Marshman:2019sne, Chen:2022wro}. In practice, one expands the interaction energy in small fluctuations about the mean separation $r_a, r_b \ll d$, and finds that the leading order coupling that induces entanglement is of the form $Gm^2 r_a r_b/d^3$. Since the oscillators were in their respective ground states to begin with, one computes the final state for the oscillator system using (first-order) perturbation theory, and show this to be an entangled state \cite{Bose:2022uxe}. As off-shell graviton degrees of freedom of get integrated out in an intermediate step, this result shows that the Newtonian potential is capable of acting as a quantum communication channel and hence entangles the two massive oscillators \cite{Christodoulou:2022mkf, Krisnanda:2019glc}. This is where quantum information theory plays a key role in understanding the implications of this experiment \cite{Belenchia:2019gcc}. As per the LOQC princtiple \cite{Bennett:1996gf}, any two quantum systems which are initially separable cannot be entangled by a local classical interaction. 

In this work, we generalize by imagining the following scenario: \textit{``What would change if the massive particles are in an expanding background?''} One motivation for doing this are the frequently encountered surprises when dealing with the vacuum state (of quantum fields) in curved spacetimes. In our thought experiment, we have two slow-moving massive, one-dimensional oscillators in an expanding universe, both of which are at rest with respect to the reference frame of the cosmological fluid (\textit{i.e.}, they move along with the Hubble flow). Even though starting out in the product state of the two ground states of the free Hamiltonians, due to the gravitational interaction between them, they end up getting entangled after a finite period of time. The precise question we seek to answer is which traits of the graviton vacuum state exhibit themselves in the resulting entanglement profile?

Instead of doing this for a general expanding universe (characterized by a FLRW spacetime), we will restrict ourselves to a de Sitter (dS) background since dS space plays a central role in our evolution history, both in the early universe (inflation) and for late-time acceleration. Moreover, this will be sufficient to illustrate our main findings. Mathematically, this means that instead of a weak field gravity regime around Minkowski used earlier, we expand gravity around a fixed dS background, where the interaction is mediated by $h_{\mu\nu}$ as before, such that the full spacetime metric is $\tilde{g}_{\mu\nu} = \eta^{\rm dS}_{\mu\nu} + h_{\mu\nu}$. The scale factor $a(t)$, which represents the relative expansion of the universe, increases exponentially with (cosmic) time, while the Hubble parameter denoted by $H \equiv \f{1}{a}\f{da}{dt}$ is a constant, for dS expansion.
\vspace{2mm}

\section{From path integrals to effective potentials}
 In generalizing from Minkowski to dS spacetime, our first job is to write down the analogue of the \textit{Newtonian potential energy} that arises due to the interaction between the oscillators mediated by the graviton field. Since there are various subtleties involved in dS space, such as not having a global time-like killing vector, it is best to use the path integral formalism to derive this in a manifestly local and (general-relativistically) covariant manner. We outline this approach \cite{Christodoulou:2022mkf} in what follows and provide some more details in the Appendix. For our system of oscillators and the gravitational field, we wish to compute the joint partition function 
\begin{equation}\label{pathintegral}
	\mc{Z} = \int \mc{D}x~\mc{DF}\exp{i\mc{S}}~,
\end{equation}
as a path integral over all particle paths $x$, and all field configurations $\mc{F}$, with the total action given by $\mc{S}$. At tree level, one can make a stationary phase approximation for the exponential functional integral in \eqref{pathintegral}, and the time-evolution operator of the system is proportional to $\exp{i\mc{S}}$ without computing the functional integral explicitly. To compute the dynamics of the oscillators in this approach, we first note that the action becomes a phase of the time evolution operator. We choose boundary conditions for the partition function such that the mediating graviton field is due to stationary masses at the initial and final particle positions. As explained in the Appendix, we are only interested in $\mc{S}_{\text{int}}$ because the phase-difference in the final state, which is the measure of the entanglement generated through the gravitational interaction,  is given by $\Delta\phi \propto \mc{S}_{\text{int}}$. The gravitational interaction potential between the two oscillators can be read off from the phase-difference as $\Delta \phi = -\int dt\, U_{\rm int}$.

While this computation is, in principle, completely general, we will invoke the non-relativistic and near-field approximations simultaneously in order to make the system analytically solvable. Through the choice of the Green's functions for the graviton field equations, we fix the Bunch-Davies vacuum state to solve for the phase-difference, and find the Newtonian limit of interaction potential to be
\begin{equation}\label{PE}
	U_{\rm int}^{\rm dS} = - \f{G~m^2}{a~\scripty{r}} - 2 Gm^2H~\ln{\left(\f{a}{a H \scripty{r}  + 1}\right)}~.
\end{equation}

\begin{figure}
	\centering
	\includegraphics[width = 1.2\textwidth]{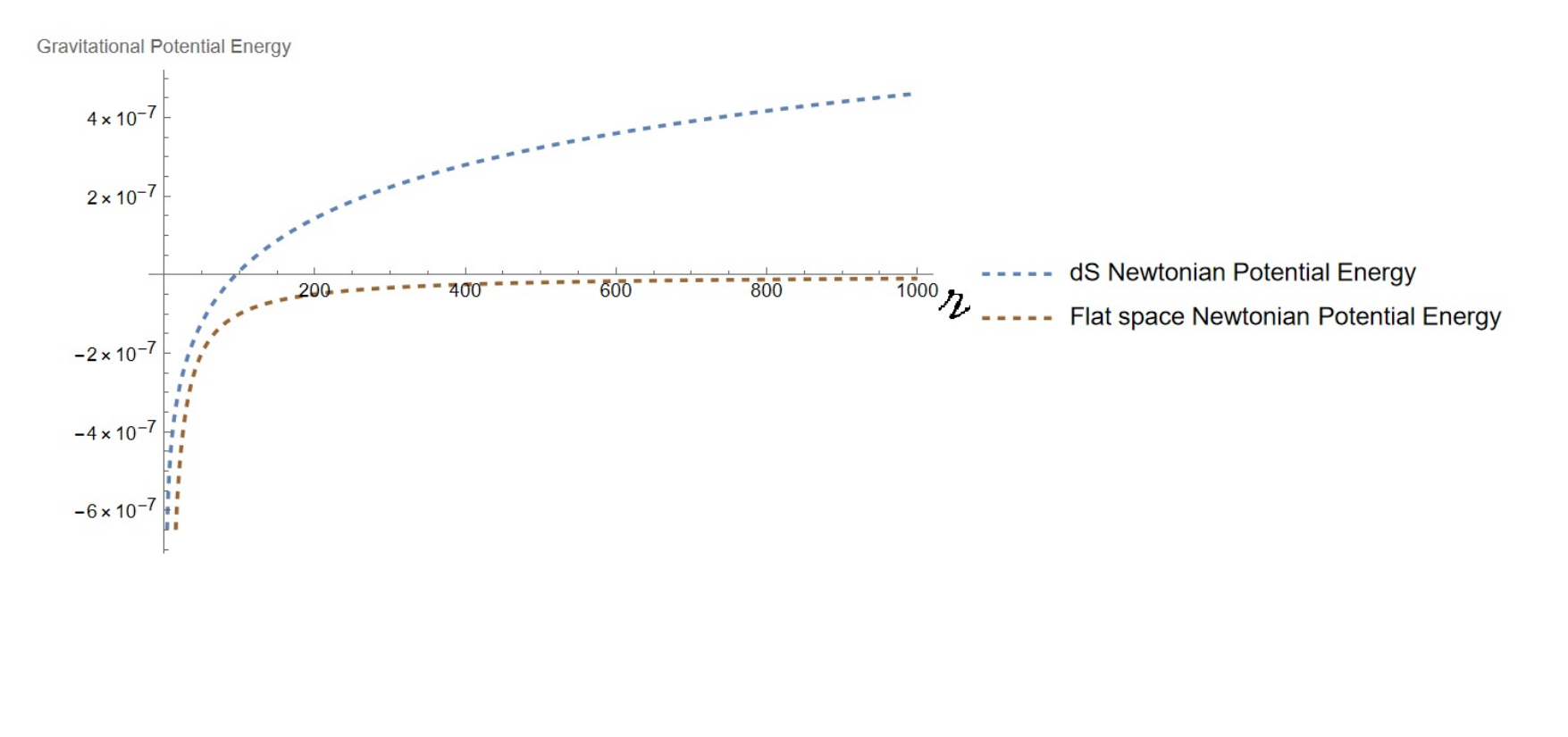}
	\caption{Comparing the Newtonian limit of the gravitational potential energy of our system in de Sitter and Minkowski backgrounds, as a function of the distance between the oscillators, as a snapshot in time (fixed by $a=10$). The Newtonian potentials are obtained by computing the path integral,  in the near-field and non-relativistic approximation, over all particle trajectories and field configurations from the on-shell action. As opposed to the usual attractive Newtonian potential in flat space ($-1/\scripty{r}$), the dS Newtonian potential turns repulsive at large separation between the masses. For simplicity, we consider the oscillators to have unit masses and take $G = 10^{-5} ({\rm GeV})^{-2}, H = 10^{-2} {\rm GeV}$, in units of $\hbar=c=1$, to enhance the effect.}
	\label{Fig01}
\end{figure}

This is our first important result. A plot of this gravitational interaction energy is shown in Fig \eqref{Fig01}. Firstly, note that this is actually an \textit{operator valued energy correction}, since the separation $\scripty{r} = |d-r_a+r_b|$ explicitly depends on the creation and annihilation operators through \eqref{QO}. What we will do is compute entanglement, due to this interaction potential, to leading order in perturbation theory. Secondly, the first term in \eqref{PE} is the usual flat-space Newtonian potential, as it should be since local physics is free from curvature effects due to the equivalence principle. Indeed, if the particles are at small separation compared to the Hubble horizon, \textit{i.e.} when $ a\scripty{r} \ll 1/H$, the potential energy reduces to (up to constant energy offsets)
\begin{equation}\label{SS}
{}^{(\rm small)} U_{\rm int}^{\rm dS} \approx - \f{G\,m^2}{a\scripty{r}} + 2\, a\, G H^2 m^2 \scripty{r} ~.
\end{equation}
At small separation, the dominance of the Newtonian potential is a universal feature of any cosmological background while the leading order correction to it is specific to dS. Since this correction term ($\propto \scripty{r} = |d + r_a -r_b|$) is of the (bi-local) form \cite{Ferrero:2021lhd}, it does not lead to any entanglement and thus, in the small separation limit, GME in dS is \textit{exactly the same} as in flat space.  However, this is a unique feature of the Bunch-Davies vacuum, implied through our choice of Green's function, used to derive the above interaction term, and hence, for an arbitrary FLRW expansion, this equivalence will not hold since the graviton mode functions will be different.

The second term in \eqref{PE} is a purely dS contribution, which dominates at large separation, \textit{i.e.} as the distance between the oscillators approaches the cosmological horizon. This term is responsible for the potential energy to change sign from being an attractive one at small distances to a repulsive one as the particle separation increases, as is expected for an accelerating universe. To see this more clearly, note that the potential energy when the oscillators are separated by distances comparable to the Hubble radius, \textit{i.e.} when $a\scripty{r} \approx 1/H$, is approximately given by  
\begin{equation}\label{LS}
	{}^{(\rm large)} U_{\rm int}^{dS} \approx 2 a\, G H^2 m^2 \left(\scripty{r}-\frac{1}{a H}\right) - \frac{5}{4}\, a^2\, G H^3 m^2\left(\scripty{r}-\frac{1}{a H}\right)^2~.
\end{equation}
Once again, the leading order term linear in $\scripty{r}$\, at $\mc{O}\left(H^2\right)$ will not lead to any entanglement, and thus, at large separations, the leading order entanglement will come from $\mc{O}\left(H^3\right)$ term. This will be the new contribution to GME between the oscillators arising solely due to dS space.
\vspace{2mm}

\section{Entanglement entropy due to GME in dS}
Although we will compute the von Neumann entanglement entropy to quantify the GME between the oscillators, any other related measure such as concurrence or log negativity would have done an equally good job. Given the Hilbert space of the two oscillators, $\mathcal{H}_a \otimes \mathcal{H}_b$, we assume that they start out in the separable vacuum state $\ket{0,0} = \ket{0}_a\otimes\ket{0}_b$. Under the action of an interaction Hamiltonian $H_{\rm int}$, corresponding to the potential \eqref{PE}, the perturbed state (to first order in perturbation theory) is given by
\begin{equation}
	\ket{\psi_f} = \f{1}{\sqrt{\mc{N}}}\sum_{n_a,n_b}\ket{n_a, n_b}~\f{\bra{n_a,n_b}{H_{\text{int}}}\ket{0,0}}{2E_0 - E_{n_a} - E_{n_b}}~.
\end{equation}
where $n_a, n_b \neq 0$. We have introduced the notation  $\ket{n_a,n_b} = \ket{n_a}\otimes \ket{n_b}$ to denote energy eigenstates of the free Hamiltonians $H^a_0$ and $H^b_0$. On tracing out one of the oscillators, one can compute the reduced density matrix of the system as $\rho_a = \sum_{n_b} \bra{n_b} \ket{\psi_f} \bra{\psi_f}\ket{n_b}$, from which one can compute the von Neumann entanglement entropy as $\sigma = -\Tr\left(\rho_a \log{\rho_a}\right)$ \cite{Balasubramanian:2011wt}. This is the standard formula for computing the entanglement entropy between two oscillators with a quantum interaction term, the role of the latter being played by gravity in dS space for our thought experiment.

The interaction Hamiltonian for some arbitrary distance, when the oscillator displacements are much smaller than their separation, \textit{i.e.} $r_a, r_b \ll d$, can be obtained by expanding \eqref{PE} as follows:
\begin{eqnarray}
    H^{\rm dS}_{\rm int} \approx \text{Non-entangling terms}  + \frac{2Gm^2}{ad^2}\left(aH - \frac{1}{d}\right)r_ar_b + \mc{O}\left(r_a^2r_b + r_ar_b^2\right) ~. 
\end{eqnarray}
Defining the coupling $\lambda_\mathfrak{g} = \f{Gm}{ad^2\omega}\left(aH - \f{1}{d}\right)$, and using (\ref{QO}), we can rewrite the interaction Hamiltonian, to leading order, as
\begin{equation}
	H_{\text{int}}^{\text{dS}} \approx  \lambda_\mathfrak{g} \left(ab+a^{\dagger}b + ab^{\dagger} + a^{\dagger}b^{\dagger}\right)~,
\end{equation}
which shows that the entanglement is generated from the $a^{\dagger}b^{\dagger}$ terms (to this order). In terms of this interaction term, the perturbed state is explicitly given by
\begin{equation}
	\ket{\psi_f} = \f{1}{\sqrt{1 + \left(\lambda_\mathfrak{g}/2\omega\right)^2}}\left\{\ket{0,0} - \f{\lambda_\mathfrak{g}}{2\omega}\ket{1,1}\right\}~, 
\end{equation}
and, therefore, the entanglement entropy is:
\begin{equation}\label{EEG}
    \sigma = -\frac{\lambda_\mathfrak{g}^2 \log \left(\frac{\lambda_\mathfrak{g}^2}{\lambda_\mathfrak{g}^2+4 \omega ^2}\right)+4 \omega ^2 \log \left(\frac{4 \omega ^2}{\lambda_\mathfrak{g}^2+4 \omega ^2}\right)}{\lambda_\mathfrak{g}^2+4 \omega ^2}~.
\end{equation}
We have plotted the behaviour of the entanglement entropy, as a function of the scale factor of the universe in Fig \eqref{EEplot}. Since the scale factor increases monotonically with respect to time for an expanding background, this plot effectively shows the time-evolution of the entanglement entropy.

\begin{figure}
    \centering
    \includegraphics{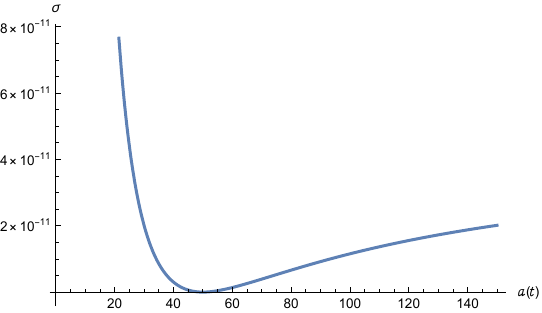}
    \caption{Entanglement entropy as a function of the scale factor from the gravitational interaction potential in dS. As the universe expands exponentially, the two oscillators get separated from each other with time (their comoving distance stays the same). Initially, the behaviour of $\sigma$ follows what is expected for flat space and falls off sharply as they get farther apart. As their separation reaches the scale of the cosmological horizon, curvature effects take over and $\sigma$ starts increasing again. This behaviour of the GME is purely due to the graviton vacuum state in dS. We consider unit masses with parameters $G = 10^{-5} ({\rm GeV})^{-2}, \omega = 0.1 {\rm GeV}, H = 10^{-2} {\rm GeV}, d = 2 {\rm GeV}^{-1}$ in units of $\hbar=c=1$, to enhance the effect.}
    \label{EEplot}
\end{figure}

\begin{figure}
    \centering 
    {{\includegraphics{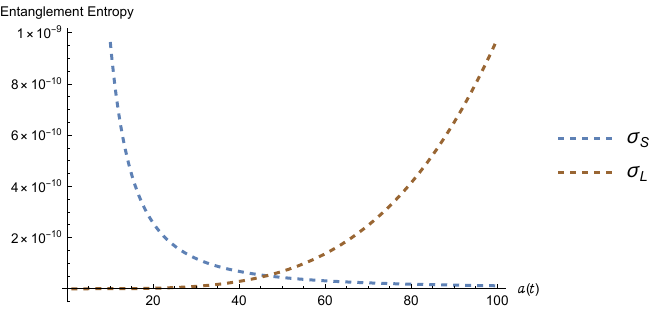} }}%
    \caption{We Taylor expand the dS Newtonian potential energy in the limits where the oscillator separation is small in one case, and when it is comparable to the horizon scale in the other, and consider the corresponding entanglement entropies ($\sigma_S, \sigma_L$ respectively) generated by the gravity-mediated interaction in these regimes. We use the same parameters as in Fig~\eqref{EEplot}.}
    \label{ComparingEE}
\end{figure}

This behaviour of the GME can be explained as follows: When the two oscillators start out very close to each other, they do not feel the effect of curvature, and the entanglement follows the one expected from Newtonian gravity in flat space. Due to accelerated expansion, the two oscillators must move farther apart from each other (in other words, although their comoving distance remains the same, their proper distance increases) and the entanglement between them falls off sharply and continues to follow the behaviour expected from flat space. However, as they approach the cosmological horizon, the unique characteristic of the background curvature shows up in the entanglement profile, and $\sigma$ starts increasing as opposed to decreasing forever as would have been the case in flat space \cite{Bose:2022uxe}. To intuitively understand this better, we have plotted the time evolution of the entanglement entropies ($\sigma_S$ and $\sigma_L$), due to the interaction Hamiltonians corresponding to the small \eqref{SS} and the large separations \eqref{LS} respectively, in Fig \eqref{ComparingEE}. Although the plots for these should not be trusted beyond their regimes of validity, collectively the two curves qualitatively show the same behaviour as exhibited by $\sigma$ for the full potential \eqref{PE}. A numerical computation shows that $\sigma_S= \sigma_L$ when $d\sim \mathcal{O}(1)/(aH)$, where $\mathcal{O}(1)$ is a numerical factor very close to $1$. Therefore, as intuitively expected, the turnover in the entanglement entropy for the full potential happens when the distance between the two oscillators is of the order of the comoving Hubble horizon of the dS universe.
\vspace{2mm}

\section{Gravity mediated entanglement in cosmology}

We have come a long way from the days of the Chappel Hill conference gendankenexperiment to propose actual GME-based table-top experiments which will prove the existence of gravitons. Nevertheless, there are still technical and conceptual gaps to fill before these experiments can be made feasible \cite{Anastopoulos:2022vvb, Grossardt:2023vpj, Toros:2020dbf}. In this article, we devise a thought experiment to see if we can improve our prospects by looking up to the skies. Firstly, it has been argued that GME experiments must operate on time-scales of the light-crossing time between the paths of the particles, maintaining spacelike separation, so as to unequivocally demonstrate the quantum nature of the interaction \cite{Martin-Martinez:2022uio}. However, such time-scales are completely out of reach of even ultra-futuristic versions of these GME-based table-top experiments. Interestingly, in our case, curvature effects show up more prominently in GME \textit{after} horizon-crossing. In other words, if we are able to detect GME in dS space in a regime where the imprints of the background curvature show up, it will necessarily demonstrate the quantum nature of gravity. One might wonder how can we ever measure entanglement if the particles go outside of the cosmological horizon. But this is not at all a far-fetched fantasy since inflationary quantum fluctuations do something very similar: they exit the horizon, classicalize and then re-enter later, which we observe as temperature anisotropies in the cosmic microwave background. Thus, it is entirely feasible to imagine that the entanglement of cosmic particles, due to their mutual gravitational interaction, will leave their signature on cosmological data if the dS-like phase is followed by decelerated cosmic expansion. Another practical problem for GME-based table-top experiments is that of decoherence due to environmental factors \cite{Aspelmeyer:2022fgc}. Once again, cosmology is a unique arena in this respect as it allows particles (such as CMB photons) to free-stream light-years across the cosmos  without interacting with anything else \cite{Berera:2021xqa}, and thus, it might be easier to avoid the problem of decoherence by looking for GME in cosmology. 

The remarkable new insight from our result \eqref{EEG} is that \textit{GME depends on background curvature for an expanding universe}.  Since the graviton mode functions for an expanding cosmos would always depend on the local expansion rate $H$, the behaviour of GME must necessarily start deviating from its flat space counterpart when the separation between the oscillators is of the order of the cosmological horizon. Even though we could explicitly demonstrate this only for dS space since it gives us an analytically solvable system, our conclusions are valid for any curved background. Thus, this line of thought would find interesting applications for particles getting entangled, due to gravity, near a black hole. Moreover, our methods go beyond gravity and would apply to other types of field-mediated entanglement in a curved background. For instance, it is possible for large primordial (electro-)magnetic fields in the early universe to entangle charged particles; however, such entanglement profiles would also get modified in a similar manner due to the effects of an expanding background. 

As fascinating as the search for existence of gravitons is, can we be even more ambitious and think of probing the spacetime curvature of the universe itself through GME? Generalizing the above thought experiment from harmonic oscillators to inflationary quantum fluctuations (which source density perturbations), we might be able to discern hidden signatures of GME in the cosmic microwave background. This, in turn, would be a complementary probe for primordial gravitational waves by putting bounds on the energy scale of inflation. On the other hand, applying GME to the current era of accelerated expansion, if observable signatures of it can be discerned in the large-scale structure data, it  will contain signals of the Hubble rate of expansion and thus give us new estimates for the age of the universe. Given the biggest crisis facing cosmology right now \cite{DiValentino:2021izs, Knox:2019rjx}, it would be amazing if (perturbative) quantum gravity can provide a fresh perspective on this problem.

\vspace{3mm}
\section*{Acknowledgements}
SB is supported in part by the Higgs Fellowship and by the STFC Consolidated Grant “Particle Physics at the Higgs Centre”. ANS is supported by an EPSRC DTP and was in part supported by SERB, DST, Government of India, under the scheme Core Research Grant (File no. CRG/2020/000616). 

\newpage

\appendix

\section*{Appendix: Deriving the gravitational interaction energy in de Sitter space}

In this section, we derive the gravitational potential energy between the two oscillators in de Sitter (dS) space from a path integral formalism. Following standard notation \cite{Christodoulou:2022mkf, Bengyat:2023hxs}, we begin by defining the path integral over all field configurations $\mc{F}$ for the graviton field, and over all trajectories $x$ of the particles, as 
\begin{equation}\label{part_func}
    \mc{Z} = \int \mc{D}x~\mc{D}\mc{F}~\exp\{i\mc{S}\}~,
\end{equation}
where $\mc{S}$ is the total action for the system. For a system of two simple harmonic oscillators being considered here, the action in the presence of gravity splits in a simple fashion as $\mc{S} = \mc{S}_0 + \mc{S}_{\text{int}}$ where $\mc{S}_{\text{int}}$ is responsible for the matter-gravity interaction and the free part of the total action $\mc{S}_0$ contains the action for the matter particles, due to the two harmonic oscillators, the Hamiltonian for which given by Eqn (1) in the main text. We wish to calculate the  (gravitational) potential energy between the two oscillators, \textit{i.e.,} we must compute the perturbation in the initial oscillator state due to the graviton field mediating the interaction.

In other words, we are essentially interested in a scattering amplitude corresponding to the exchange of off-shell gravitons between the two oscillators \textit{at tree-level}. Therefore, we use the stationary-phase approximation to evaluate the partition function \eqref{part_func}. The time-evolution operator turns the total action into a phase; moreover, the $\mc{S}_0$ part will only contribute to an overall global phase in the partition function \cite{Christodoulou:2022mkf} which will be the same for both the oscillators. Therefore, we will be only concerned with  the $\mc{S}_{\text{int}}$ part of the action since we are only interested about the phase-difference between the paths of the two oscillators in the final (perturbed) state of the system.

Furthermore, note that there is a contribution to the total action coming from the expansion of the gravitational action to quadratic order, which we have not written in $\mc{S}$ above. Recall that we are linearizing gravity around a dS space in a weak-field regime, \textit{i.e.} $\tilde{g}_{\mu\nu} = \eta^{\rm dS}_{\mu\nu} + h_{\mu\nu}$, where the dS metric is given by $\eta^{\rm dS}_{\mu\nu} = \frac{1}{H^2\tau^2} \left(-d\tau^2 +d\textbf{x}^2\right)$. The scale factor is given by $a(\tau)=-1/(H \tau)$, where $H$ is the Hubble parameter and $\tau$ is the conformal time coordinate running from $(-\infty, 0)$. However, since it has been shown that the phase difference in the final state of the system comes from the \textit{on-shell} action  in the stationary phase approximation \cite{Christodoulou:2022mkf}, we do not need to even write down the quadratic action of the gravitational field $h_{\mu\nu}$, as this will vanish on-shell. Thus, we can proceed to calculate $\mc{S}_{\text{int}}$ given by the usual term
\begin{eqnarray}\label{Sint}
    \mc{S}_{\text{int}} = \int d^4x \sqrt{-g}~ h_{\mu\nu}T^{\mu\nu}~,
\end{eqnarray}
where $T_{\mu\nu}$ is the stress energy tensor of the two oscillator system given below and we assume that they are of equal mass:
\begin{equation}
    T^{\mu\nu} =  \sum_{\mathfrak{e}}\f{m}{\sqrt{-g}}\f{ds_{\mathfrak{e}}}{d\tau}\left[\delta^3(x^i - r^i_{\mathfrak{e}})\f{d\chi^{\mu}_{\mathfrak{e}}}{ds_{\mathfrak{e}}}\f{d\chi^{\nu}_{\mathfrak{e}}}{ds_{\mathfrak{e}}}\right]~,
\end{equation}
where $\mathfrak{e} = a,b$ stands for the two oscillators, $s_{\mathfrak{e}}$ is proper time of oscillator $\mathfrak{e}$,   $ds_{\mathfrak{e}}/d\tau = a/\gamma_{\mathfrak{e}}$, with $\gamma_{\mathfrak{e}} = \sqrt{1-v^2_{\mathfrak{e}}}$. Note that we have defined the notation $\chi^{\mu}_{\mathfrak{e}} = \left(\tau_{\mathfrak{e}}, r_{\mathfrak{e}}^i\right)$ to denote the four-position of the particles.  Imposing the slow-moving (non-relativistic) approximation will be akin to taking the limit $\gamma_{\mathfrak{e}} \to 1$. Sticking to linearized gravity means that indices will be raised and lowered using $\eta^{\rm dS}_{\mu\nu}$. 

We first wish to calculate $h_{\mu\nu}$ by solving the (linearized) field equations 
\begin{eqnarray}
   \Box h_{\mu\nu} = - 16\pi~G~\bar{T}_{\mu\nu}~, ~~~~~~~~~~{\rm where} ~~~\bar{T}_{\mu\nu} = T_{\mu\nu} - \f{1}{2}\eta^{\rm dS}_{\mu\nu}\eta^{{\rm dS}~\alpha\beta}\,T_{\alpha\beta}~,
\end{eqnarray}
by using the method of Green's functions. The boundary conditions used for deriving the Green's functions automatically fixes the graviton state to be that of the Bunch Davies vacuum (through the choice of the corresponding mode functions), and we ignore the backreaction of the particles on the graviton state. The relevant retarded Green's function is given by \cite{Glavan:2019yfc}
\begin{equation}
    G_R(x,x') = -\f{\theta(\Delta\tau)}{2\pi}\left[\f{\delta(\Delta\tau^2 - |\Delta \textbf{x}|^2)}{a(\tau)a(\tau')} + \f{H^2}{2}\theta(\Delta\tau - |\Delta \textbf{x}|)\right]~,
\end{equation}
where  $x \equiv (\tau,\textbf{x})$. We have defined $\Delta\textbf{x} = \textbf{x} - \textbf{x}'$ and $\Delta \tau = \tau - \tau'$ with the source term given explicitly by
\begin{eqnarray}
    \bar{T}_{\mu\nu} &=& g_{\mu\alpha}g_{\nu\beta}\bar{T}^{\alpha \beta} = \sum_{\mathfrak{e}}\underbrace{g_{\mu\alpha}g_{\nu\beta}\left\{\f{d\chi^{\mu}_{\mathfrak{e}}}{ds_{\mathfrak{e}}}\f{d\chi^{\nu}_{\mathfrak{e}}}{ds_{\mathfrak{e}}} + \f{1}{2}g^{\mu\nu}\right\}}_{M_{\mu\nu\,{\mathfrak{e}}}} \f{m~\delta^3(x^i - r^i_{\mathfrak{e}})}{a^3~\gamma_{\mathfrak{e}}}\nonumber\\
    &\equiv& \sum_{\mathfrak{e}}M_{\mu\nu\,{\mathfrak{e}}}\f{m~\delta^3(x^i - r^i_{\mathfrak{e}})}{a^3~\gamma_{\mathfrak{e}}}~.
\end{eqnarray}
Therefore, the graviton field solution is evaluated as
\begin{align}\label{formalh}
    h_{\mu\nu}(x^i,\tau) &= -4\pi~G\int d^3x'~d\tau'~a^4(\tau')G_R(x,x')~\bar{T}_{\mu\nu}~\nonumber \\ 
     &= 2G\int d^3x'~\int_{\tau_0}^{0}d\tau'~a^4(\tau')\,\theta(\Delta\tau)\left[\f{\delta(\Delta\tau^2 - |\Delta \textbf{x}|^2)}{a(\tau)a(\tau')} + \f{H^2}{2}\theta(\Delta\tau - |\Delta \textbf{x}|)\right]\times\nonumber\\
    &~~~~~~~~~~~~~~~~~~~~~~~~~~~~~~~~~~~\sum_{\mathfrak{e}}M_{\mu\nu\,{\mathfrak{e}}}\f{m~\delta^3(x'^i - r^i_{\mathfrak{e}})}{a(\tau')^3~\gamma_{\mathfrak{e}}(\tau')}~.
\end{align}

We carry out the integration as follows. Firstly, note that one can decompose the integral over the $\delta$ function in $\tau'^2$ into a sum of two $\delta$ function integrals, each linear in $\tau'$ using the property, 
\begin{equation*}
    \delta(f(y)) = \sum_i\f{\delta(y - y_i)}{\vert f'(y_i) \vert}
\end{equation*}
where $y_i$ are the roots of the equation $f(y) = 0$. This gives us:
\begin{align}
    h_{\mu\nu}(x^i,\tau) & = 2G\sum_{\mathfrak{e}}m_{\mathfrak{e}}\left\{\int_{\tau_0}^0d\tau'\f{1}{\gamma_{\mathfrak{e}}(\tau')~a(\tau)}\left[\f{\delta(\tau' - \tau_+)}{2r_{\mathfrak{e}}} + \f{\delta(\tau'-\tau_-)}{2r_{\mathfrak{e}}}\right]~M_{\mu\nu~{\mathfrak{e}}}\right\} \nonumber \\
    & ~~~~~~~~~~~~~~~~~~~~+\left\{\int_{\tau_0}^{\tau - r_{\mathfrak{e}}}d\tau' \,\f{a(\tau')~H^2}{2}~M_{\mu\nu~{\mathfrak{e}}}\right\}~,
\end{align}
with $\tau_- = \tau - |\textbf{x}^i - \chi^i_{\mathfrak{e}}|$ and $\tau_+ = \tau + |\textbf{x}^i - \chi^i_{\mathfrak{e}}|$ being the roots of the polynomial $\Delta\tau^2 - \vert\Delta\textbf{x}\vert^2 = 0$. Recognising that $\tau_-$ is the retarded time, and that the other root does not contribute to the integral given the limits of the integral, we can discard this latter time to naturally get the on-shell field solution in terms of the retarded time $\tau_R$, which is implicitly defined by $\tau - \tau_R = \vert r_b - r_a\vert$. 

Plugging this into \eqref{Sint} and taking the slow moving approximation, we evaluate the on-shell action as: 
\begin{align}
    \mc{S}_{\text{int}} &\approx G~m^2\int d\tau~\int d^3x~\delta^3(x^i - \chi^i_b)~\left\{\f{1}{r_a}+ a(\tau)\int^{\tau-r_a}_{\tau_0}d\tau'~\left(-\f{H}{\tau'}\right)\right\}~.
\end{align}
Note that we have also used the explicit time dependence of the scale factor in dS as $a(\tau) = -1/H\tau$ in the second integral above.

By definition, the effective interaction potential can be derived from the on-shell action as follows:
\begin{equation}
    \mc{S}_{\text{int}} = -\int~dt~U_{\rm int}~.
\end{equation}
Therefore, we can carry out the spatial integration and rewrite the integral over the conformal time as an integral over the cosmic time through the scale factor ($a d\tau=dt$). Finally, we arrive at the effective potential
\begin{equation}
    U^{\rm non\, rel}_{\rm int}(\scripty{r}) = -{G~m^2}\left\{\f{1}{a~\scripty{r}~(\tau_R)} + H\log{\left(\f{a}{aH\scripty{r}~(\tau_R) + 1}\right)}\right\}~,
\end{equation}
where we have not yet employed the near field approximation (but have imposed the slow-moving one). Invoking the near field approximation is equivalent to replacing the retarded time with the coordinate (conformal) time and is essentially the Newtonian limit of interaction potential in dS. This is what we quote in Eqn. (8) in the main draft.


\begin{thebibliography}{100}
\bibitem{Chapman:2021jbh}
S.~Chapman and G.~Policastro, ``{Quantum computational complexity from quantum
	information to black holes and back},'' \emph{Eur. Phys. J. C}, vol.~82,
no.~2, p. 128, 2022.

\bibitem{DeWitt:1957obj}
C.~M. De~Witt, Ed., \emph{{Proceedings: Conference on the Role of Gravitation
		in Physics, Chapel Hill, North Carolina, Jan 18-23, 1957}}, 1957.

\bibitem{Mari:2015qva}
A.~Mari, G.~De~Palma, and V.~Giovannetti, ``{Experiments testing macroscopic
	quantum superpositions must be slow},'' \emph{Sci. Rep.}, vol.~6, p. 22777,
2016.

\bibitem{Dyson:2014qoa}
F.~Dyson, ``{Is a graviton detectable?}'' in \emph{{Conference in Honor of the
		90th Birthday of Freeman Dyson}}, 2014, pp. 1--14.

\bibitem{Parikh:2020nrd}
M.~Parikh, F.~Wilczek, and G.~Zahariade, ``{The Noise of Gravitons},''
\emph{Int. J. Mod. Phys. D}, vol.~29, no.~14, p. 2042001, 2020.

\bibitem{Bose:2017nin}
S.~Bose, A.~Mazumdar, G.~W. Morley, H.~Ulbricht, M.~Toro\v{s}, M.~Paternostro,
A.~Geraci, P.~Barker, M.~S. Kim, and G.~Milburn, ``{Spin Entanglement Witness
	for Quantum Gravity},'' \emph{Phys. Rev. Lett.}, vol. 119, no.~24, p. 240401,
2017.

\bibitem{Marletto:2017kzi}
C.~Marletto and V.~Vedral, ``{Gravitationally-induced entanglement between two
	massive particles is sufficient evidence of quantum effects in gravity},''
\emph{Phys. Rev. Lett.}, vol. 119, no.~24, p. 240402, 2017.

\bibitem{Belenchia:2018szb}
A.~Belenchia, R.~M. Wald, F.~Giacomini, E.~Castro-Ruiz, v.~Brukner, and
M.~Aspelmeyer, ``{Quantum Superposition of Massive Objects and the
	Quantization of Gravity},'' \emph{Phys. Rev. D}, vol.~98, no.~12, p. 126009,
2018.

\bibitem{Marletto:2018lsb}
C.~Marletto and V.~Vedral, ``{When can gravity path-entangle two spatially
	superposed masses?}'' \emph{Phys. Rev. D}, vol.~98, p. 046001, 2018.

\bibitem{Bose:2022uxe}
S.~Bose, A.~Mazumdar, M.~Schut, and M.~Toro\v{s}, ``{Mechanism for the quantum
	natured gravitons to entangle masses},'' \emph{Phys. Rev. D}, vol. 105,
no.~10, p. 106028, 2022.

\bibitem{Anastopoulos:2015zta}
C.~Anastopoulos and B.-L. Hu, ``{Probing a Gravitational Cat State},''
\emph{Class. Quant. Grav.}, vol.~32, no.~16, p. 165022, 2015.

\bibitem{Danielson:2021egj}
D.~L. Danielson, G.~Satishchandran, and R.~M. Wald, ``{Gravitationally mediated
	entanglement: Newtonian field versus gravitons},'' \emph{Phys. Rev. D}, vol.
105, no.~8, p. 086001, 2022.

\bibitem{Biswas:2022qto}
D.~Biswas, S.~Bose, A.~Mazumdar, and M.~Toro\v{s}, ``{Gravitational
	optomechanics: Photon-matter entanglement via graviton exchange},''
\emph{Phys. Rev. D}, vol. 108, no.~6, p. 064023, 2023.

\bibitem{Christodoulou:2018cmk}
M.~Christodoulou and C.~Rovelli, ``{On the possibility of laboratory evidence
	for quantum superposition of geometries},'' \emph{Phys. Lett. B}, vol. 792,
pp. 64--68, 2019.

\bibitem{Danielson:2022tdw}
D.~L. Danielson, G.~Satishchandran, and R.~M. Wald, ``{Black holes decohere
	quantum superpositions},'' \emph{Int. J. Mod. Phys. D}, vol.~31, no.~14, p.
2241003, 2022.

\bibitem{Christodoulou:2022knr}
M.~Christodoulou, A.~Di~Biagio, R.~Howl, and C.~Rovelli, ``{Gravity
	entanglement, quantum reference systems, degrees of freedom},'' \emph{Class.
	Quant. Grav.}, vol.~40, no.~4, p. 047001, 2023.

\bibitem{doi:10.1142/S0218271823420208}
S.~Brahma and A.~N. Seenivasan, ``{Gravity-induced entanglement as a probe of
	spacetime curvature},'' \emph{Int. J. Mod. Phys. D}, vol.~0, no.~0, p.
2342020, 2023.

\bibitem{Yant:2023smr}
J.~Yant and M.~Blencowe, ``{Gravitationally induced entanglement in a harmonic
	trap},'' \emph{Phys. Rev. D}, vol. 107, no.~10, p. 106018, 2023.

\bibitem{Marshman:2019sne}
R.~J. Marshman, A.~Mazumdar, and S.~Bose, ``{Locality and entanglement in
	table-top testing of the quantum nature of linearized gravity},'' \emph{Phys.
	Rev. A}, vol. 101, no.~5, p. 052110, 2020.

\bibitem{PhysRevLett.127.123602}
J.~Hinney, A.~S. Prasad, S.~Mahmoodian, K.~Hammerer, A.~Rauschenbeutel,
P.~Schneeweiss, J.~Volz, and M.~Schemmer, ``Unraveling two-photon
entanglement via the squeezing spectrum of light traveling through
nanofiber-coupled atoms,'' \emph{Phys. Rev. Lett.}, vol. 127, p. 123602, Sep
2021.

\bibitem{PhysRevLett.130.240203}
Z.~Mehdi, J.~J. Hope, and S.~A. Haine, ``Signatures of quantum gravity in the
gravitational self-interaction of photons,'' \emph{Phys. Rev. Lett.}, vol.
130, p. 240203, Jun 2023.

\bibitem{Christodoulou:2022mkf}
M.~Christodoulou, A.~Di~Biagio, M.~Aspelmeyer, v.~Brukner, C.~Rovelli, and
R.~Howl, ``{Locally Mediated Entanglement in Linearized Quantum Gravity},''
\emph{Phys. Rev. Lett.}, vol. 130, no.~10, p. 100202, 2023.

\bibitem{Chen:2022wro}
L.-Q. Chen, F.~Giacomini, and C.~Rovelli, ``{Quantum States of Fields for
	Quantum Split Sources},'' \emph{Quantum}, vol.~7, p. 958, 2023.

\bibitem{Krisnanda:2019glc}
T.~Krisnanda, G.~Y. Tham, M.~Paternostro, and T.~Paterek, ``{Observable quantum
	entanglement due to gravity},'' \emph{npj Quantum Inf.}, vol.~6, p.~12, 2020.

\bibitem{Belenchia:2019gcc}
A.~Belenchia, R.~M. Wald, F.~Giacomini, E.~Castro-Ruiz, v.~Brukner, and
M.~Aspelmeyer, ``{Information Content of the Gravitational Field of a Quantum
	Superposition},'' \emph{Int. J. Mod. Phys. D}, vol.~28, no.~14, p. 1943001,
2019.

\bibitem{Bennett:1996gf}
C.~H. Bennett, D.~P. DiVincenzo, J.~A. Smolin, and W.~K. Wootters, ``{Mixed
	state entanglement and quantum error correction},'' \emph{Phys. Rev. A},
vol.~54, pp. 3824--3851, 1996.

\bibitem{Ferrero:2021lhd}
R.~Ferrero and C.~Ripken, ``{De Sitter scattering amplitudes in the Born
	approximation},'' \emph{SciPost Phys.}, vol.~13, p. 106, 2022.

\bibitem{Balasubramanian:2011wt}
V.~Balasubramanian, M.~B. McDermott, and M.~Van~Raamsdonk, ``{Momentum-space
	entanglement and renormalization in quantum field theory},'' \emph{Phys. Rev.
	D}, vol.~86, p. 045014, 2012.

\bibitem{Anastopoulos:2022vvb}
C.~Anastopoulos and B.-L. Hu, ``{Gravity, Quantum Fields and Quantum
	Information: Problems with Classical Channel and Stochastic Theories},''
\emph{Entropy}, vol.~24, no.~4, p. 490, 2022.

\bibitem{Grossardt:2023vpj}
A.~Gro\ss{}ardt and M.~Kemal~D\"oner, ``{Lessons and complications from
	gravitationally induced entanglement},'' \emph{J. Phys. Conf. Ser.}, vol.
2533, no.~1, p. 012022, 2023.

\bibitem{Toros:2020dbf}
M.~Toro\v{s}, T.~W. Van De~Kamp, R.~J. Marshman, M.~S. Kim, A.~Mazumdar, and
S.~Bose, ``{Relative acceleration noise mitigation for nanocrystal
	matter-wave interferometry: Applications to entangling masses via quantum
	gravity},'' \emph{Phys. Rev. Res.}, vol.~3, no.~2, p. 023178, 2021.

\bibitem{Martin-Martinez:2022uio}
E.~Mart\'\i{}n-Mart\'\i{}nez and T.~R. Perche, ``{What gravity mediated
	entanglement can really tell us about quantum gravity},'' 8 2022.

\bibitem{Aspelmeyer:2022fgc}
M.~Aspelmeyer, ``{When Zeh Meets Feynman: How to~Avoid the~Appearance
	of~a~Classical World in~Gravity Experiments},'' \emph{Fundam. Theor. Phys.},
vol. 204, pp. 85--95, 2022.

\bibitem{Berera:2021xqa}
A.~Berera, S.~Brahma, R.~Brandenberger, J.~Calder\'on-Figueroa, and A.~Heavens,
``{Quantum coherence of photons to cosmological distances},'' \emph{Phys.
	Rev. D}, vol. 104, no.~6, p. 063519, 2021.

\bibitem{DiValentino:2021izs}
E.~Di~Valentino, O.~Mena, S.~Pan, L.~Visinelli, W.~Yang, A.~Melchiorri, D.~F.
Mota, A.~G. Riess, and J.~Silk, ``{In the realm of the Hubble
	tension\textemdash{}a review of solutions},'' \emph{Class. Quant. Grav.},
vol.~38, no.~15, p. 153001, 2021.

\bibitem{Knox:2019rjx}
L.~Knox and M.~Millea, ``{Hubble constant hunter\textquoteright{}s guide},''
\emph{Phys. Rev. D}, vol. 101, no.~4, p. 043533, 2020.

\bibitem{Bengyat:2023hxs}
O.~Bengyat, A.~Di~Biagio, M.~Aspelmeyer, and M.~Christodoulou, ``{Gravity
	Mediated Entanglement between Oscillators as Quantum Superposition of
	Geometries},'' 9 2023.

\bibitem{Glavan:2019yfc}
D.~Glavan, S.~P. Miao, T.~Prokopec, and R.~P. Woodard, ``{Breaking of scaling
	symmetry by massless scalar on de Sitter},'' \emph{Phys. Lett. B}, vol. 798,
p. 134944, 2019.

\end{thebibliography}
\end{document}